\documentclass[11pt,a4paper]{article}

\usepackage{authblk}
\usepackage{hyperref}
\usepackage{graphicx}
\usepackage{pdfpages}

\begin{document}

\title{SIMLR: a tool for large-scale genomic analyses by multi-kernel learning}
 
\author[1]{Bo Wang\footnote{Equal contributors. Correspondance to \href{bowang87@stanford.edu}{bowang87@stanford.edu} or \href{daniele.ramazzotti@stanford.edu}{daniele.ramazzotti@stanford.edu}.}}
\author[1,2]{Daniele Ramazzotti$^{\ast}$}
\author[3]{Luca De Sano}
\author[4]{Junjie Zhu}
\author[1]{Emma Pierson}
\author[1]{Serafim Batzoglou}
\affil[1]{Department of Computer Science, Stanford University, Stanford, CA, USA}
\affil[2]{Department of Pathology, Stanford University, Stanford, CA, USA}
\affil[3]{Department of Informatics, University of Milano-Bicocca, Milan, Italy}
\affil[4]{Department of Electrical Engineering, Stanford University, Stanford, CA, USA}

\date{}

\maketitle

\begin{abstract}
\textbf{Motivation:} We here present SIMLR (\underline{S}ingle-cell \underline{I}nterpretation via \underline{M}ulti-kernel \underline{L}ea\underline{R}ning), an open-source tool that implements a novel framework to learn a sample-to-sample similarity measure from expression data observed for heterogenous samples. SIMLR can be effectively used to perform tasks such as dimension reduction, clustering, and visualization of heterogeneous populations of samples. SIMLR was benchmarked against state-of-the-art methods for these three tasks on several public datasets, showing it to be scalable and capable of greatly improving clustering performance, as well as providing valuable insights by making the data more interpretable via better a visualization. \\
\textbf{Availability and Implementation:} SIMLR is available on \href{https://github.com/BatzoglouLabSU/SIMLR}{GitHub} in both R and MATLAB implementations. Furthermore, it is also available as an R package on \href{http://bioconductor.org/}{bioconductor.org}. 
\end{abstract}

The recent development of high resolution single-cell RNA-seq (scRNA-seq) technologies increases the availability of high throughput gene expression measurements of individual cells. This allows us to dissect previously unknown heterogeneity and functional diversity among cell populations \cite{shapiro2013single}. In this line of work recent efforts (see \cite{pollen2014low,usoskin2015unbiased,kolodziejczyk2015single}) have demonstrated that \textit{de novo} cell type discovery of functionally distinct cell sub-populations is possible via unbiased analysis of all transcriptomic information provided by scRNA-seq data. However, such analysis heavily relies on the accurate assessment of pairwise cell-to-cell similarities, which poses unique challenges such as outlier cell populations, transcript amplification noise, and dropout events (\textit{i.e.,} zero expression measurements due to sampling or stochastic transcriptional activities) \cite{pierson2015zifa}. 

Recently, new single-cell platforms such as DropSeq \cite{macosko2015highly} and GemCode single-cell technology \cite{zheng2016massively} have enabled a dramatic increase in throughput to hundreds of thousands of cells. While such technological advances may add additional power for \textit{de novo} discovery of cell populations, they also increase computational burdens for traditional unsupervised learning methods. 

To address all of the aforementioned challenges, SIMLR was originally proposed in \cite{wang2017visualization} as a novel framework capable of learning an appropriate cell-to-cell similarity metric from the input single-cell data. However, although originally proposed for the analysis of single-cell data, SIMLR can be effectively adopted in the broader task of studying biological data describing heterogeneous populations including but not limited to single-cell analysis (see below and the Supplemenatary Materials). The learned similarities in fact can be exploited for multible tasks, such as effective dimension reduction, clustering, and visualization. SIMLR provides a more scalable analytical framework, which works on hundreds of thousands of samples without any loss of accuracy in dissecting heterogeneity. 

\vspace{0.5cm}

SIMLR is available in both R and MATLAB implementations. The framework is capable of learning similarities among gene expression data within an heterogeneous populations of samples, which have been shown to capture different representations of the data. To this end, the approach combines multiple Gaussian kernels in an optimization framework, which can be efficiently solved by a simple iterative procedure. Moreover, SIMLR addresses the challenge of high levels of noise and dropout events by employing a rank constraint and graph diffusion in the learned similarity \cite{wang2014similarity}. See Figure $1$ for an overview of the framework. 

\begin{figure}[!ht]
\centering
\includegraphics[width=1.00\textwidth]{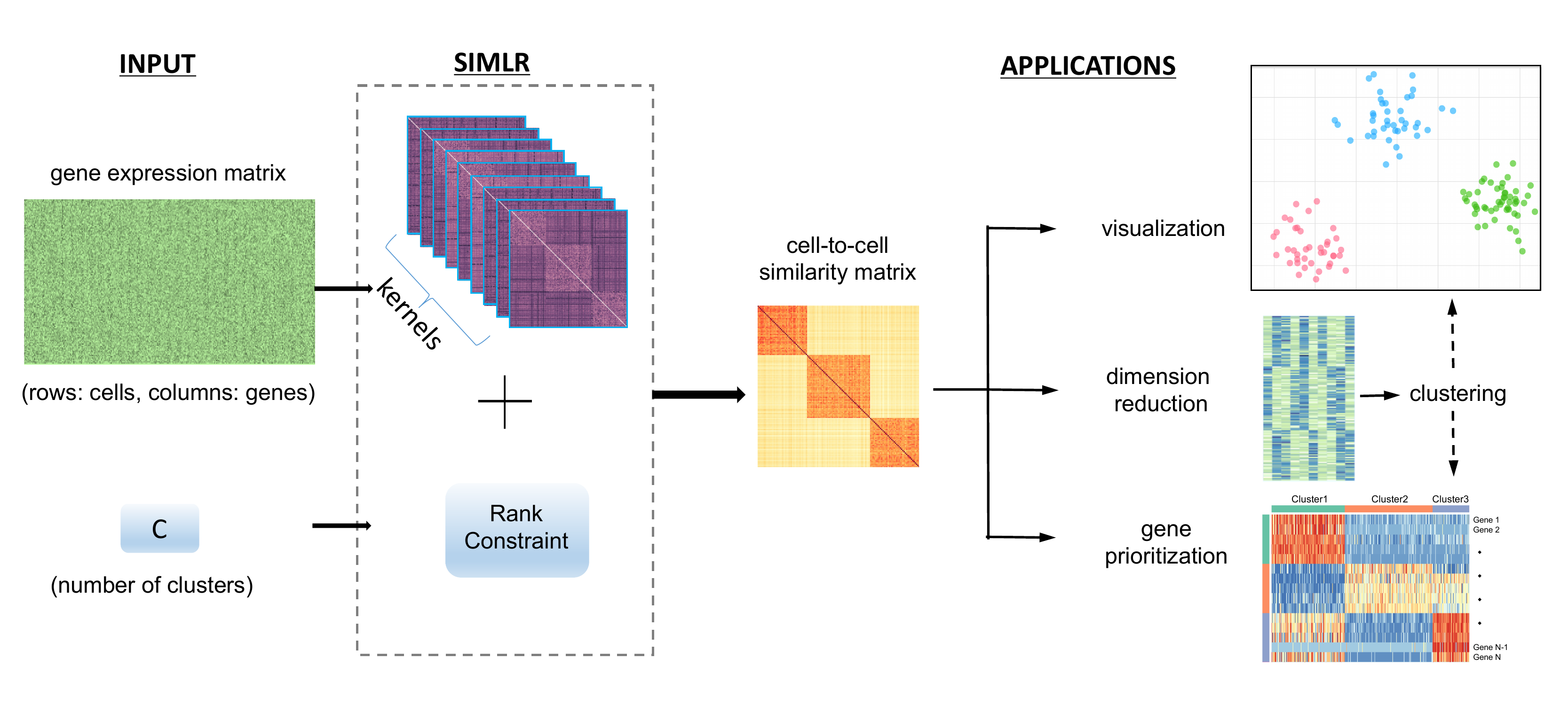}
\caption[SIMLR pipeline \cite{wang2014similarity}.]{We start with an input matrix with gene expression observations for a set of genes. SIMLR is then capable of learning a set of sample-to-sample similarities by estimating multiple kernels, with the assumptions of the presence of $C$ separable populations within the data. To this extent, SIMLR constraints the similarity matrix to have an approximate block-diagonal structure with $C$ blocks where the samples of the same populations to be more similar. The learned similarities can be used for multiple tasks; they can be used for visualization, reduce the dimension of the data, cluster the populations into subgroups and prioritize the most variable genes that explain the differences across the populations.}
\label{fig:simlr_framework}
\end{figure}

In the tool are provided both a standard implementation and a large-scale extension of SIMLR together with two examples to test the methods on the datasets by \cite{buettner2015computational} for the standard SIMLR and \cite{zeisel2015cell} for the large-scale extension (see Supplementary Material for details). SIMLR can accurately analyze both datasets within minutes on a single core laptop. 

Moreover, we also report in the Supplementary Materials a complementary example of usage of SIMLR to study heterogeneity in a cancer dataset. Specifically, we consider the data from \cite{cancer2015comprehensive} and we report how our framework can also be applied in this context, first by estimating the number of populations as discussed in \cite{wang2014similarity} and then by learning a patient-to-patient similarity which may allow, e.g., to effectively stratify the tumors. 

An other advantage of SIMLR is that the learned similarities can be efficiently adopted for multiple downstream applications. Some applications include prioritizing genes by ranking their concordance with the similarity and creating low-dimensional representations of the samples by transforming the input into a stochastic neighbor embedding framework, all of which is implemented in our software. We refer to the Supplementary Material for detailed use cases of the tool and to \cite{wang2017visualization} for a detailed description of the method and for several applications on genomic data from public datasets. 

\vspace{0.5cm}

In conclusions, SIMLR can infer similarities that can be used to perform dimension reduction, clustering, and visualization in different contexts, with the goal of better understandying the underlying heterogeneity of the studied phenomenon. While the multiple-kernel learning framework has obvious advantages on heterogeneous datasets, where several clusters coexist, we also believe that this approach, together with its visualization framework, may also be valuable for data that does not contain clear clusters, such as cell populations that contain cells spanning a continuum or a developmental pathway. 

\bibliographystyle{unsrt}
\bibliography{bibliography}



\section*{Supplementary material}

# SIMLR: a tool for large-scale genomic analyses by multi-kernel learning

*Supplementary Information*

Bo Wang*  Daniele Ramazzotti*  Luca De Sano
Junjie Zhu  Emma Pierson  Serafim Batzoglou

SIMLR is available in R and Matlab. We will now present details about each implementation together with examples of analyses on 2 single-cell datasets and 1 NGS cancer dataset.

# 1 R

## 1.1 Installation

The *SIMLR R package* is available on *Bioconductor* at `https://bioconductor.org/packages/release/bioc/html/SIMLR.html` and can be installed as follows.

```
## try http:// if https:// URLs are not supported
source("https://bioconductor.org/biocLite.R")
biocLite("SIMLR")
```

The package is also available on *Github* at the address `https://github.com/BatzoglouLabSU/SIMLR`. It is possible to install both the master (stable) and development versions of the R package by using the R library *devtools*.

```
library(devtools)
install_github("BatzoglouLabSU/SIMLR", ref = 'master')
library(SIMLR)

library(devtools)
install_github("BatzoglouLabSU/SIMLR", ref = 'development')
library(SIMLR)
```

## 1.2 Examples on Single-cell data

We now show two use cases for SIMLR in order to highlight the main features of our tool. We first load the data provided as an example in the package. The

---

*Equal contributors.

dataset *BuettnerFlorian* [1] is used to illustrate how to use standard SIMLR, while a reduced version of the dataset *ZeiselAmit* [2] is used to illustrate how to use large-scale SIMLR.

```
library(SIMLR)
data(BuettnerFlorian)
data(ZeiselAmit)
```

The external R package *igraph* [3] is required for the computation of the normalized mutual information to assess the results of the clustering.

```
library(igraph)
```

We run SIMLR on the BuettnerFlorian input dataset. For this dataset we have a ground truth of 3 cell populations (clusters).

```
set.seed(11111)
example = SIMLR(X = BuettnerFlorian$in_X,
                c = BuettnerFlorian$n_clust)
```

```
## Computing the multiple Kernels.
## Performing network diffiusion.
## Iteration:  1
## Iteration:  2
...
## Iteration:  10
## Iteration:  11
## Performing t-SNE.
## Epoch: Iteration # 100  error is:  0.1140084
## Epoch: Iteration # 200  error is:  0.06181848
...
## Epoch: Iteration # 900  error is:  0.05889082
## Epoch: Iteration # 1000  error is:  0.0588387
## Performing Kmeans.
## Performing t-SNE.
## Epoch: Iteration # 100  error is:  10.36092
## Epoch: Iteration # 200  error is:  1.167142
...
## Epoch: Iteration # 900  error is:  0.793667
## Epoch: Iteration # 1000  error is:  0.6030175
```

To assess the performance of our method, we compute the normalized mutual information (NMI) between the clusters inferred by SIMLR and the ground truth clusters. NMI takes values between 0 and 1, with higher values reflecting better performance.

```
nmi = compare(BuettnerFlorian$true_labs[,1], example$y$cluster,
              method="nmi")
```

```r
print(nmi)
```

```
## [1] 0.888298
```

To visualize the results, we plot the cell populations in Figure 1.

```r
plot(example$ydata,
  col = c(topo.colors(BuettnerFlorian$n_clust))[BuettnerFlorian$true_labs[,1]],
  xlab = "SIMLR component 1",
  ylab = "SIMLR component 2",
  pch = 20,
  main="SIMILR 2D visualization for BuettnerFlorian")
```

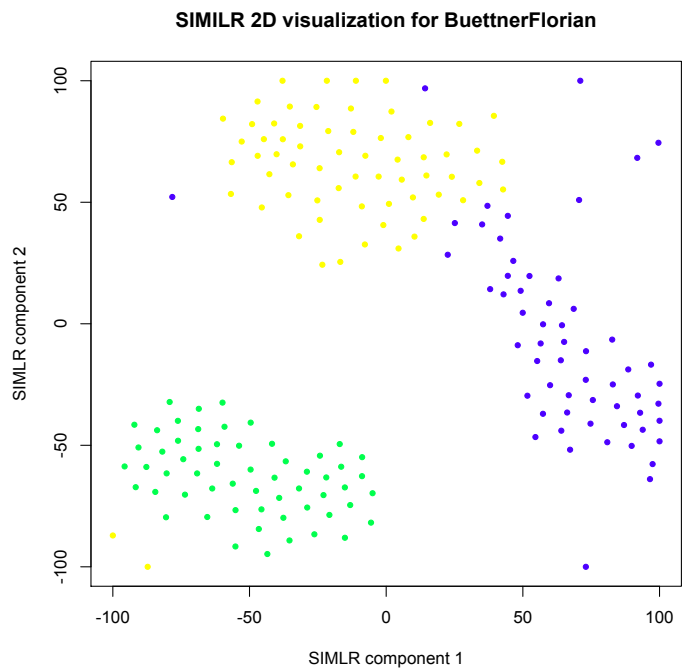

Figure 1: Example of visualization by SIMLR for the BuettnerFlorian dataset.

We also run SIMLR feature ranking on the same inputs to get a rank of the key genes with the related pvalues.

```r
set.seed(11111)
ranks = SIMLR_Feature_Ranking(A=BuettnerFlorian$results$S,
                              X=BuettnerFlorian$in_X)
head(ranks$pval)
```

```
## [1] 2.201015e-125  2.531379e-90  5.632172e-77  6.719501e-76
##       4.444251e-72  8.822900e-69
```
```r
head(ranks$aggR)
```
```
## [1] 5701 1689 7549   57 2653 8081
```

We now provide an example application of large-scale SIMLR to an input dataset (a reduced version of the dataset provided in Zeisel, Amit, et al). The full dataset has 9 cell populations, but for the sake of this example, we use a reduced version with only 2 clusters.

```r
set.seed(11111)
example_large = SIMLR_Large_Scale(X = ZeiselAmit$in_X,
                                  c = ZeiselAmit$n_clust)
```
```
## Performing fast PCA.
## Performing k-nearest neighbour search.
## Computing the multiple Kernels.
## Performing the iterative procedure  5  times.
## Iteration:  1
## Iteration:  2
## Iteration:  3
## Iteration:  4
## Iteration:  5
## Performing Kmeans.
## Performing t-SNE.
## The main loop will be now performed with a maximum of
## 300 iterations.
## Performing iteration 1.
## Performing iteration 2.
...
## Performing iteration 299.
## Performing iteration 300.
```

Once again, we evaluate the performance of SIMLR by computing the NMI between large-scale SIMLR's inferred clusters and the ground truth clusters.

```r
nmi_large = compare(ZeiselAmit$true_labs[,1],example_large$y$cluster,
                    method="nmi")
print(nmi_large)
```

```
## [1] 0.9348853
```

We plot the cell populations in Figure 2.

```r
plot(example_large$ydata,
  col = c(topo.colors(ZeiselAmit$n_clust))[ZeiselAmit$true_labs[,1]],
  xlab = "SIMLR component 1",
  ylab = "SIMLR component 2",
  pch = 20,
  main="SIMLR 2D visualization for ZeiselAmit")
```

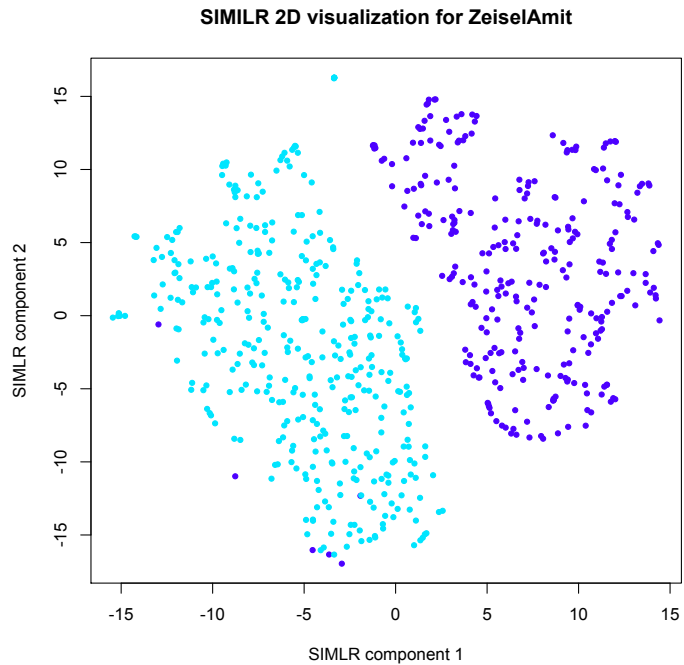

Figure 2: Example of visualization by SIMLR for the ZeiselAmit dataset.

## 2 Matlab

### 2.1 Installation

The Matlab version of SIMLR is available on *Github* at `https://github.com/BatzoglouLabSU/SIMLR` (SIMLR branch). The tool may be installed as follows.

```
%%% Installation
% Before running SIMLR, the user needs to mex compile several C-mex files.
% You can simply run the code INSTALL_SIMLR.m

INSTALL_SIMLR.m
```

### 2.2 Examples on Single-cell data

We now provide code to run the Matlab implementation of SIMLR on two examples: one for standard SIMLR and one for the large-scale implementation. The code and datasets are available on *Github*.

```
%%% For small-scale Single-cell RNA-seq with less than 3000 cells,
%%% please use SIMLR.m
```

```matlab
% Given the gene expression matrix in_X, (if in_X is the gene counts
% matrix, please use log10(1+in_X) first), we can run SIMLR as follows:

[y, S, F, ydata] = SIMLR(in_X,C,K);

%%% Inputs are as follows:
% in_X is the input gene expression matrix of size NxM, where N is
% the number of cells and M is the number of genes.
% C is the number of clusters,
% K is the number of neighbors, and by default, K =10

%%% The outputs are as follows:
% y is the obtained labels
% S is the learned similarity of size NxN, where N is the number of cells
% F is the latent variables of size NxC
% ydata is the 2-D visualization of cells

%%% You can plot the visualization as follows:
scatter(ydata(:,1),ydata(:,2));

% or if you have any labels, you can color-code the cells and show the
% visualization as follows:
SIMLR_DisplayVisualization(ydata,true_labs);

%%% Once you have your similarity S, you can run feature selection as follows:
aggR = SIMLR_Feature_Ranking(S,in_X);

% aggR is a vector of ranking for M genes. Usually we take the top 100 genes as
% the most important/differential genes.

%%% For large-scale single-cell RNA-seq data with more than 3000 cells,
%%% we recommend using SIMLR_LARGE.m as follows:

% Step 0: If the input is gene counts, we take log10 transformations:
in_X = log10(1+in_X);

% Step 1: Learn the similarity S and the latent embedding F
[S, F] = SIMLR_LARGE(in_X,C,K); % K is usually set to be 30~50 for large scale

% Step 2: Running clustering on F:
y = litekmeans(F,C,'Replicates',50);

% Step 3: Running visualization from S:
% d is the dimension for the visualization, with a default value of 2
ydata = SIMLR_embedding_tsne(S,1,d,F(:,1:2));
```

```
% Step 3.1: Show your visualization
SIMLR_DisplayVisualization(ydata,true_labs)
```

## 3 Analysis of NGS cancer data

We now show the results of applying our tool on NGS cancer data as the ones provided by TCGA studies [4]. Specifically, we refer to the cohort of patients of [5] and we consider as input, expression data for a total of 282 patients and 20890 genes. The input data is provided (gliomas.RData for R and gliomas.mat for Matlab are provided in the Github repository of the tool).

We show in the next Sections the steps of this analysis in the Matlab implementation of SIMLR.

### 3.1 Estimating the number of clusters

As a first step of out analysis, we estimate the best number of clusters for these data with the heuristics discussed in [6] as follow.

```
% load the data
load('gliomas.mat')

% estimate the best number of clusters
rng(4302992);
NUMC = 2:15;
[K1, K2] = Estimate_Number_of_Clusters_SIMLR(gliomas, NUMC);
```

The results of this analysis are shown in Figure 3 and consist of 2 heuristics evaluated for a set of possible number of clusters (shown on the left and on the right in the figure for number of clusters from 2 to 15). We choose the best number of clusters by picking the lower point for both of the 2 metrics, i.e., in this case 12.

### 3.2 Clusters by SIMLR and visualization

We now perform the standard SIMLR analysis and ask for 12 clusters.

```
% estimate the best number of clusters
rng(5492003);
C = 12;
[y, S, F, ydata] = SIMLR(my_data,C,10);
```

As an assesment to explore the results, we show in Figure 4 the visualization of the 12 clusters which results in beeing very clearly separated.

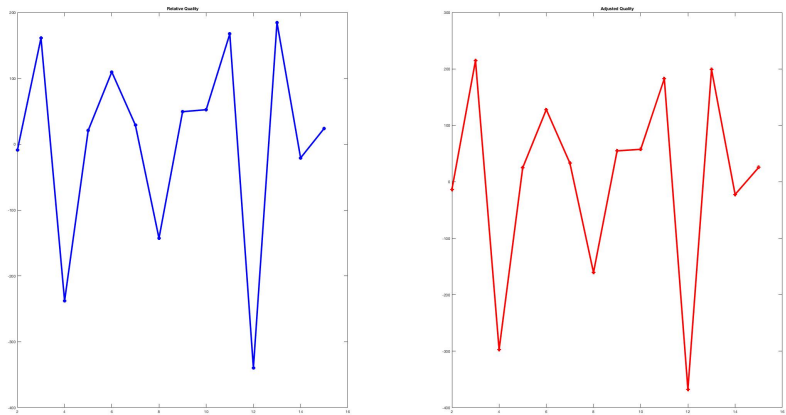

Figure 3: Example of estimation of the best number of clusters by SIMLR for the lower grade gliomas dataset of [5].

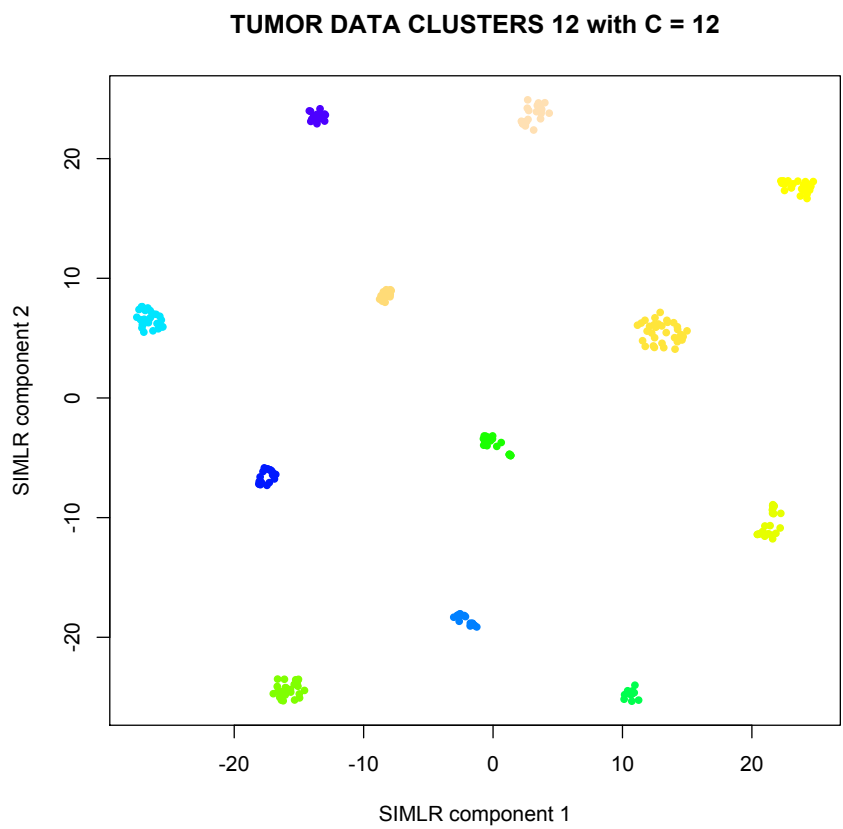

Figure 4: Visualization by SIMLR for the the lower grade gliomas dataset of [5].



\end{document}